\documentclass[seceq]{ptptex}
\usepackage[dvips]{graphicx}
\usepackage{amsmath}

\title{New class of symmetries for self-gravitating hydrodynamics equations}

\author{Souichi \textsc{Murata} \footnote{Corresponding author. Tel: +81-52-789-3553; Fax: +81-52-789-2906. \\E-mail address: smurata@allegro.phys.nagoya-u.ac.jp } and Kazuhiro Nozaki}

\inst{Department of Physics, Nagoya University, Chikusa-ku, Nagoya, 464-8602, Japan}

\abst{A method of calculating a new class of symmetries is presented for partial differential equations. The method give a new dynamical solution for an isothermal and  cylindrically symmetric hydrodynamics equations under self-gravity. The solution  describes a development of a gas cylinder from a motion-less state to  states of expansion and oscillation.}

\begin{document}
\maketitle

\section{Introduction}

\ Constructing exact solutions to a system of partial differential equations (PDEs) is an important problem in mathematical physics.  One of the most powerful methods to find  particular solutions to a system of PDEs is based upon the analysis of its invariance with respect to the Lie group \cite{Olver,Kumei,CRC}.  If a system of PDEs is invariant under a Lie group of point transformation, we can construct special solutions, called similarity solutions, which are invariant under the group  admitted by the system.  This method provides a systematic and unified procedure in search of similarity solutions.

However, many systems of non-linear PDEs in mathematical physics admit only trivial symmetry groups, for example, translational symmetries with respect to time and space, rotational symmetries and scaling symmetries and so on. Invariant solutions under such trivial symmetry groups are also constructed by other naive methods based on physical considerations.  If we find a method to derive a new non-trivial symmetry, a new type of self-similarity solutions may be constructed by means of the symmetry.

In this paper, we present a method to obtain a new class of symmetry groups admitted by a system of PDEs. A key of our treatment is to carry out symmetry group analysis on particular solutions of physical interest.  To illustrate our method,  we apply the method to the  equations of cylindrically symmetric isothermal hydrodynamics under self-gravity \cite{Inutsuka} . A Similarity solution for the self-gravitating isothermal gaseous cylinder without rotation has been discussed by Inutsuka and Miyama \cite{Inutsuka} . In their study, they gave a self-similarity solution for a collapsing or expanding isothermal filament whose density follows the Emden-type equation with an inertial term.  Here, we study dynamics of the filament in the presence of a rotational flow around the axis of the cylinder by mean of our method of Lie group and find a new class of symmetries and a new quasi-self similarity solution with a rotational flow.

The plan of the paper is as follow: in section 2, we explain our method to find a  new class of symmetries to a system of PDEs. In section 3, we apply our method to investigation of  a quasi-similarity solution in the presence of a  rotational flow. 

\section{Mathematical description of symmetry group analysis}

A system of n-th order PDEs in $p$ independent and $q$ dependent variables is given as a system of equations

\begin{align}
\Delta_{\nu}(\mathbf{x},\mathbf{u}^{(n)})=0, \quad \nu= 1 \dots l, 
                                                            \label{PDEs}
\end{align}  
involving $\mathbf{x}=(x^1, \dots, x^p)$, $\mathbf{u}=(u^1,\dots,u^q)$ and the derivatives of $\mathbf{u}$ with respect to $\mathbf{x}$ up to n, where $\mathbf{u}^{(n)}$ represents all the derivatives of $\mathbf{u}$ of all orders from 0 to n.  

We consider a one-parameter Lie group of infinitesimal transformations acting on the independent and dependent variables of the system

\begin{align}
\tilde{x}^i &=x^i+\epsilon \xi^i(\mathbf{x},\mathbf{u})+O(\epsilon^2),
                                                       \quad i=1,\dots,p, \\
\tilde{u}^j &=u^j+\epsilon \Phi^j(\mathbf{x},\mathbf{u})+O(\epsilon^2),
                                                       \quad j=1,\dots,q. 
\end{align}
The infinitesimal generator $\mathbf{V}$ associated with the above group of  transformations can be written as 

\begin{align}
\mathbf{V}=\sum^{p}_i \xi^i(\mathbf{x},\mathbf{u})\partial_{x^i}
                +\sum^{q}_j \Phi^j(\mathbf{x},\mathbf{u})\partial_{u^j}.
                                                        \label{generator} 
\end{align}
The Invariance of the system (\ref{PDEs}) under the infinitesimal transformations leads to the invariance condition

\begin{align}
Pr^{(n)}\mathbf{V}[\Delta_{\nu}(\mathbf{x},\mathbf{u}^{(n)})]=0,
                          \quad \nu=1,\dots l, \ \ \mbox{whenever} \ \  \Delta_{\mu}(\mathbf{x},\mathbf{u}^{(n)})=0, \label{prolongation}
\end{align}
where $Pr^{(n)}$ is the n-th order prolongation of the infinitesimal generator.

Since the coefficients of the infinitesimal generator do not include derivatives of $\mathbf{u}$,  we can separate (\ref{prolongation}) with respect to derivatives and solve the resulting overdeterminied system of linear homogeneous PDEs known as the determining equations.  Our purpose for calculating symmetries is to obtain similarity solutions which are invariant under the symmetries. Applying the criterion of invariance, we get the invariant conditions called the functional self-similarity conditions

\begin{align}
\mathbf{V}(u^i-\overline{u}^i)\bigg|_{u^i=\overline{u}^i}=0, \label{V}
\end{align}
where $\overline{u}^j$ is the solution of the system.  In order to solve the conditions (\ref{V}), we have to integrate the Lie equation corresponding to $\mathbf{V}$ 

\begin{align}
 \frac{dx^1}{\xi^1}= \dots = \frac{dx^m}{\xi^m}
=\frac{du^1}{\Phi^1}=\dots = \frac{du^n}{\Phi^n}.  \label{Lie-eq}
\end{align}
Then, Eq (\ref{V}) and solutions of (\ref{Lie-eq}) yield an invariant (self-similarity) solution under the symmetry.

The crucial point is how to find a new non-trivial symmetry of the given equations.  Our approach is to restrict solutions on some special class, for example, in the following application we consider the Emden-type solution where the pressure is nearly balanced with the gravity. The restriction is given as an additional system of PDEs in general

\begin{align}
\tilde{\Delta}(\mathbf{x},\mathbf{u}^{(n)})=S(\mathbf{x},\mathbf{u}),
                                                     \label{PDEs-add}
\end{align}
where $S(\mathbf{x},\mathbf{u})$ is a function of $\mathbf{x}$ and $\mathbf{u}$
and will be  given in calculating the determining equations. Then, the number of the determining equations is reduced owing to addition of the auxiliary equation (\ref{PDEs-add}) so that a new class of symmetries may emerge. To demonstrate how we can construct a non-trivial symmetry and a new solution,  we apply the our method to the self-gravitating hydrodynamics equations.

\section{Self-gravitating hydrodynamics for polytropic gas cylinder}

The self-gravitating hydrodynamics equations of a polytropic gas is often introduced in studies of the different stages of stellar evolution \cite{Chand} . It often happens that the dynamics of a self-gravitating system exhibits self-similar behavior. A new similarity solution obtained either analytically or numerically, although constituting only a limited part of the general physical solution, may be very useful in understanding the basic dynamical behavior of the system. 

An isothermal polytropic gas cylinder is described by the following normalized fluid equation.

\begin{eqnarray}
\rho_t+u\rho_r+\rho \bigg(u_r+\frac{u}{r}\bigg)=0,\label{cont}\\
   u_t+u u_r -\frac{v^2}{r}+\frac{1}{\rho}\rho_r+\sigma_r=0,\label{motion-r}\\
   v_t+u v_r+\frac{1}{r}uv=0,\label{motion-angular}\\
\sigma_{rr}+\frac{\sigma_r}{r}=\rho, \label{Poisson}
\end{eqnarray}
where the pressure $p$ , density $\rho$ , radial velocity  $u$ ,  angular velocity $v$ , gravitational potential $\sigma$ and radius $ r$ are normalized by 
$p_0, \rho_0,u_0=v_0=(\gamma p_0/\rho_0)^{1/2},\gamma p_0/\rho_0,  r_0, $  respectively; $G$ is the gravitational constant and $p=\rho^\gamma$. The acoustic time scale $ r_0/u_0,$ where $ r_0=\gamma p_0/(4\pi G \rho_0^2)$ , is set to equal to the time scale of free fall $(4\pi G\rho_0)^{-1/2}$.

Since we are interested in the Emden-type solution where the pressure is nearly balanced with self-gravity,  we impose the following additional condition

\begin{align}
\sigma_r+\frac{1}{\rho_r}\rho_r=-S(r,t),  \nonumber
\intertext{or}
 u_t+u u_r -\frac{v^2}{r}=S(r,t), \label{add}
\end{align}
where $S(r,t)$ is a function of $r$ and $t$.
 
Let us seek a Lie symmetry admitted by the system of Eqs.(\ref{cont}), (\ref{motion-r}), (\ref{motion-angular}), (\ref{Poisson}) and (\ref{add}). The function $S(r,t)$ should be given by the determining equations for the symmetry. Applying Pr$^{(1)}\mathbf{V}$ to Eqs.(\ref{cont}), (\ref{motion-r}), (\ref{motion-angular}), (\ref{Poisson}) and  (\ref{add}), we obtain a dozen of determining equations for the symmetry. After some straightfoward calculations of the determining equations, we obtain the following symmetry.

\begin{align}
\mathbf{V}=\tau(t)\partial_t+\frac{1}{2}\tau_tr\partial_r
             -\tau_t \rho\partial_{\rho}
             +\frac{1}{2}(-\tau_t u+\tau_{tt}r)\partial_u
   -\frac{1}{2}\tau_t v\partial_v -\frac{1}{2}\tau_t \phi\partial_{\phi},
                                    \label{symmetry}  \\
\tau_{ttt}\tau-2a\tau_t=0.   \label{tau}\\
S(r,t)=\frac{\tau_{ttt}}{2\tau_t}r, \label{S}  
\end{align}
where $\phi \equiv \sigma_r$ and $a$ is an arbitrary constant.  The invariance under the symmetry group (\ref{symmetry}) leads to the conditions of functional self-similarity on Eqs. (\ref{cont}), (\ref{motion-r}), (\ref{motion-angular}), (\ref{Poisson})

\begin{align}
-\tau_t \rho-\frac{1}{2}\tau_t r \rho_r
           +\tau \bigg( u\rho_r+\rho \bigg(u_r+\frac{u}{r}\bigg) \bigg)=0,
                                                      \label{FSS-rho}\\
\frac{1}{2}(-\tau_t u+\tau_{tt}r)-\frac{1}{2}\tau_t r u_r
  +\tau \bigg( u_r -\frac{v^2}{r}+\frac{1}{\rho}\rho_r+\phi \bigg)=0,
                                                      \label{FSS-u}\\
 -\frac{1}{2}\tau_t v-\frac{1}{2}\tau_t r v_r
          +\tau \bigg( u v_r+\frac{1}{r}uv \bigg)=0,  \label{FSS-v}\\
-\frac{1}{2}\tau_t \phi -\frac{1}{2}\tau_t r \bigg(\rho-\frac{1}{r}\phi \bigg)
          -\tau \phi_t=0,   \label{FSS-phi}.
\end{align}
Let us choose the radial velocity as 

\begin{align}
u=\frac{\tau_t}{2\tau}r, \label{u}
\end{align}
so that Eqs. (\ref{FSS-rho}) and (\ref{FSS-v}) are satisfied automatically. Then, Eqs (\ref{add}), (\ref{S}), (\ref{u}) give

\begin{align}
v^2=\bigg[ \bigg( \frac{\tau_t}{2\tau} \bigg)_t+ 
           \bigg( \frac{\tau_t}{2\tau} \bigg)^2-
           \frac{\tau_{ttt}}{2\tau_t}         \bigg] r^2.  \label{consistent}
\end{align}
It is easy to see that Eq. (\ref{tau}) takes the following integrable form

\begin{align}
f f_{ttt}+3f_t f_{tt}-2a \frac{f_t}{f}=0, \label{f} 
                           \qquad  \mbox{where} \ \ \  f(t)=\sqrt{\tau}, 
\intertext{which can be integrated once and yields}
f_{tt}=-\frac{d V}{df},   \qquad V \equiv \frac{b}{2f^2}-a\log(f), 
                                                      \label{potential}
\end{align}
where $b$ is an integral constant.

Solving the Lie equation corresponding to the infinitesimal generator (\ref{symmetry}), we get a new solution 

\begin{align}
y=\frac{r}{f},\quad \rho=\frac{R(y)}{f^2},\quad u=f_t y,
             \quad v^2=b\frac{y^2}{f^2}, \quad \phi=\frac{\Phi(y)}{f}, 
                                                              \label{solution}
\intertext{where the integral constant $b$ should be positive. According to Eqs.(\ref{FSS-u}) and (\ref{FSS-phi}), the function $R(y)$ and $\Phi(y)$ are expressed as the solution of the Emden-type equation}
\frac{d^2 \theta}{d y^2}+\frac{1}{y}\frac{d \theta}{d y}
         +\exp (\theta)+2a=0, \quad   \Phi=-\frac{d \theta}{dy}-ay, 
          \quad \theta \equiv\log (R),\label{Emden}
\end{align}
where a term $2a$ comes from the inertia.

\begin{figure}[t]
\begin{center}
\rotatebox{-90}
{\includegraphics[width=6.0cm]{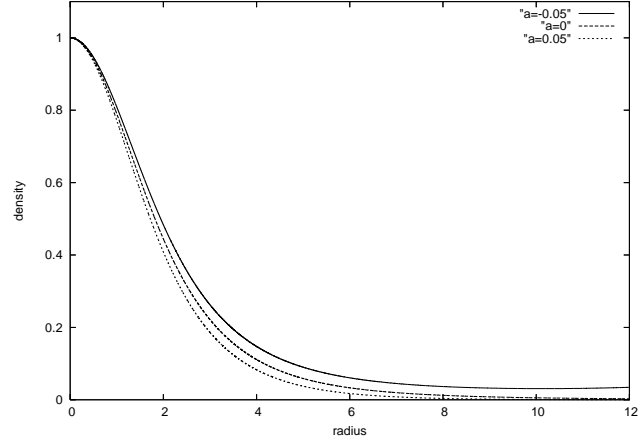}}
\caption{density profile for $a=0$, $a=0.05$ and $a=-0.05$ with the boundary condition $R(0)=1$, $R_y(0)=0$ }
\end{center}
\end{figure}

For $a=b=0$, the symmetry (\ref{symmetry}) becomes a naive Lie point symmetry admitted by Eqs.(\ref{cont}), (\ref{motion-r}), (\ref{motion-angular}), (\ref{Poisson}) and the solution (\ref{solution}) coincides with a trivial self-similarity solution. For $a \neq 0$ and $b=0$, the solution (\ref{solution}) corresponds to the Inutsuka-Miyama solution. For other values of $a$ and $b$ ,the solutiion represents  a new quasi-similarity solution.

In Fig.1, typical solutions of the Emden-type equation (\ref{Emden}) are plotted in the case  $a=0$, $a=-0.05$ and  $a=0.05$. A constant $y$ , i.e., $y=r/f(t)=y_0$ gives a radially moving frame, through which the gas dose not flow since   $u|_{y=y_0}=dr/dt|_{y=y_0}$. So, we can  put the boundary of the density (the surface of the gas cylinder)  on $y=y_0$ . The dynamics of the boundary depends on the evolution of $f(t)$ only, which is determined by the potential $V(f)$ given in Eq.(\ref{potential})

Let us  consider an initial value problem for the Eq. (\ref{f}) such that $f(t)|_{t=0}=1$ in order that $y|_{t=0}=r$. The evolution of $f(t)$ is classified into three cases.

First of all, we consider the case  $a \geq 0$ and $b \geq 0$. In this case, we can take a motionless state initially so that $f_t|_{t=0}=0$ and $f_{tt}|_{t=0}=a+b$. Then, the surface is accelerated due to the positive inertia force. Radial and angular components of the velocity at the surface are depicted for $a=1$ and $b=0.1,0.5,1.0$ in Fig.2 and 3.

\begin{figure}[t]
\begin{center}
\rotatebox{-90}
{\includegraphics[width=6.0cm]{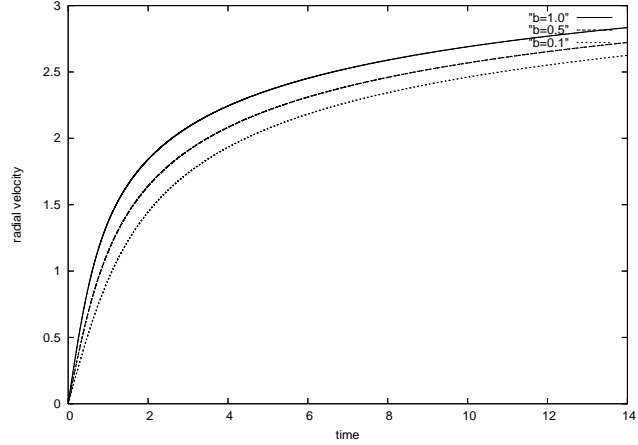}}
\caption{Radial velocity for $b=0.1$, $b=0.5$ and $b=1.0$ .}
\end{center}
\end{figure}

\begin{figure}[h]
\begin{center}
\rotatebox{-90}
{\includegraphics[width=6.0cm]{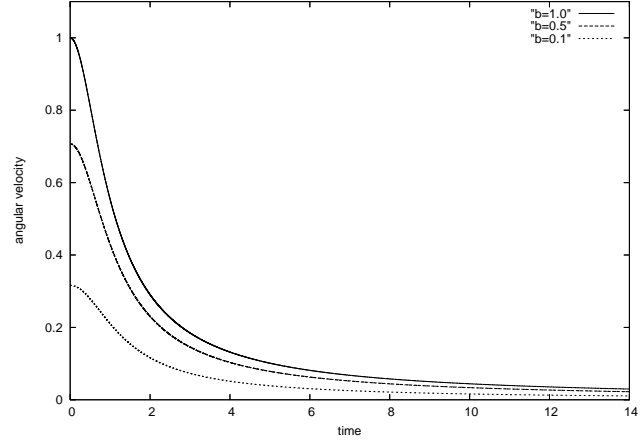}}
\caption{Angular velocity for $b=0.1$, $b=0.5$ and $b=1.0$ .}
\end{center}
\end{figure}

Next we consider a radially decelerating case without a rotational flow  (i.e $a<0$ and $b=0$), where we take $f_t|_{t=0}=v_0 (>0)$ as an initial condition so that the initial velocity of the boundary is positive. Then an initially expanding surface gradually decelerates due to the negative inertia term: Eq.(\ref{potential}) can be integrated once and reads

\begin{align}
\frac{1}{2}\bigg( \frac{df}{dt} \bigg)^2+V(f)=E_0.   \label{energy}
\end{align}
A reflection point $f_r$ appears at 

\begin{align}
f=f_r=\exp\bigg\{ \frac{v_0^2}{2|a|} \bigg\}.  \nonumber
\end{align}
The surface stops at the reflection point and then the gas cylinder began to shrinks. Reflection time $t_r$ is given as 

\begin{align}
t_r=\int_1^{f_r}\frac{df}{\sqrt{2(E_0-|a|\log (f))}}. \nonumber
\end{align}
In this reflection case, the radial velocity of boundary is depicted for $a=-0.5$ and $v_0=1$ in Fig.4, where $t_r=4.06$.

\begin{figure}[t]
\begin{center}
\rotatebox{-90}
{\includegraphics[width=6.0cm]{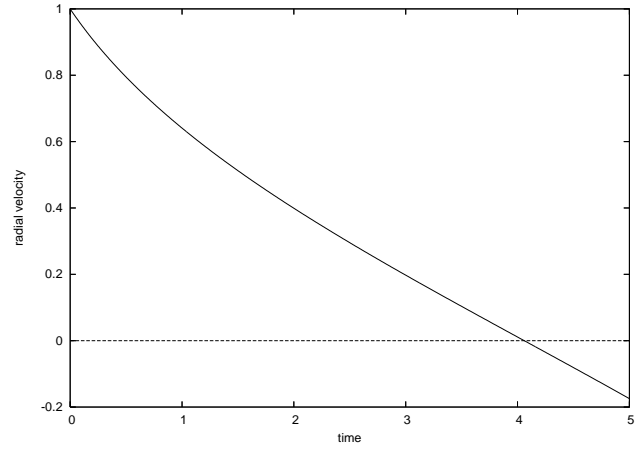}}
\caption{Radial velocity of the boundary for $a=-0.5$.}
\end{center}
\end{figure}

Finally, we consider an oscillation case $a<0$ and $b > 0$. The potential $V(f)$ has a concave form and has an equilibrium point at $f=f_e=\sqrt{b/|a|}$.  Let us estimate a frequency around the equilibrium point,  we expand $f$ for small $\xi$ 

\begin{align}
f&=\sqrt{\frac{b}{|a|}}+\xi,\qquad  \mbox{where} \  \xi \ll 1,
\intertext{Substituting this expansion into Eq.(\ref{energy}), we have}
\frac{1}{2} \bigg(\frac{df}{dt} \bigg)^2 &\approx 
       E_0 -\bigg\{
                 V_e+\frac{d V}{df}\bigg|_e \xi 
                    +\frac{1}{2}\frac{d^2 V}{df^2}\bigg|_e \xi^2 
            \bigg\}    \nonumber \\
      &=\epsilon_0-b\bigg( \frac{|a|}{b} \bigg)^2\xi^2, \label{expantion}
\end{align}
where the suffixation $e$ of the each term denotes one at the equilibrium point and $\epsilon_0=E_0-V(\sqrt{b/|a|})$. Therefore, the displacement $\xi$ is given as a harmonic oscillation $\xi=A \sin (\omega t)$ where

\begin{align}
A=\sqrt{\frac{b \epsilon_0}{a^2}},  \qquad \omega=\sqrt{\frac{2 a^2}{b}} 
                                                                   \nonumber
\end{align}
In Fig.5,6 and 7, we show numerically calculated oscillating patterns  $f$, the radial velocity and the angular velocity for $a=-1.0$, $b=2.0$ and $v_0=0.0$. For this values of the parameters, the period is 7.23, which is larger than $2\pi/ \omega$.

\begin{figure}[t]
\begin{center}
\rotatebox{-90}
{\includegraphics[width=6.0cm]{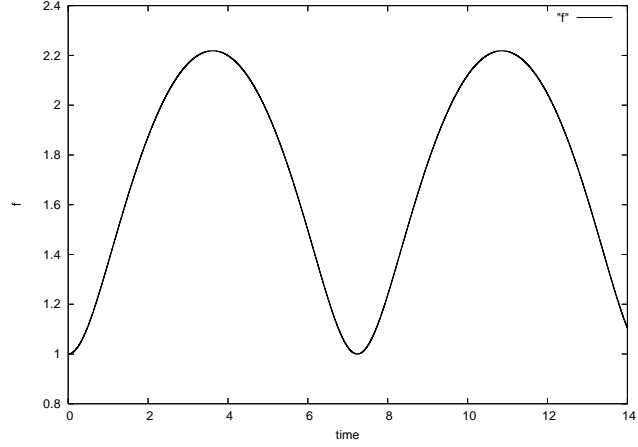}}
\caption{Radius $f$ for $a=-1.0$, $b=2.0$ and  $v_0=0$ .}
\end{center}
\end{figure}

\begin{figure}[h]
\begin{center}
\rotatebox{-90}
{\includegraphics[width=6.0cm]{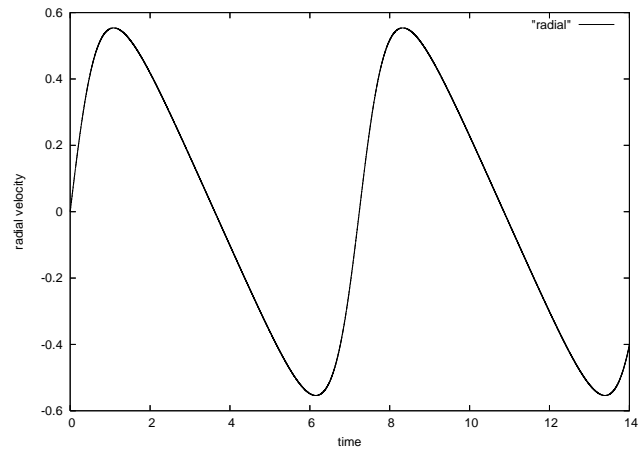}}
\caption{Radial velocity for $a=-1.0$, $b=2.0$ and  $v_0=0$ .}
\end{center}
\end{figure}

\begin{figure}[h]
\begin{center}
\rotatebox{-90}
{\includegraphics[width=6.0cm]{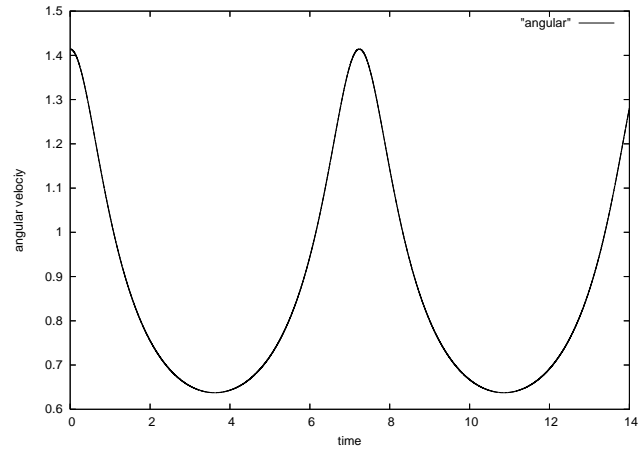}}
\caption{Angular velocity for $a=-1.0$, $b=2.0$ and  $v_0=0$ .}
\end{center}
\end{figure}

\section{summary}

We present a new method to the Lie symmetry analysis for non-linear PDEs.  The key of our method is to calculate the Lie symmetry on special solutions. In mathematical physics, we often seek special solutions describing some physical phenomena.  Characterizing a special class of solutions by an additional system, we find a broader class of symmetries.

To illustrate our method, we consider the Emden-type solution in self-gravitating hydrodynamics for an isothermal gas cylinder in the presence of rotating flow and find a new class quasi-similarity solution.

The solution is classified by two parameters $a$ and $b$, namely, the accelerating case ($a \geq 0$ and $b \geq 0$), the decelerating case ($a < 0$ and $b=0$) and the oscillating case ($a < 0$ and $ b >0$). The frequency near the equilibrium point in the third case is given by the values of parameters $a$ and $b$.

\end{document}